\documentstyle[11pt, epsf]{article}

\title{\bf Cumulative particle production and percolation of strings}
\author{M.A. Braun \thanks{On leave of absence from:
 Department of High Energy physics,
 University of S.Petersburg,
198904 S.Petersburg, Russia}, E.G.Ferreiro,
F. del Moral and C. Pajares\\
Department of Particle Physics\\
University of Santiago de Compostela\\
15706 Santiago de Compostela, Spain}
\def\beq{\begin{equation}}
\def\eeq{\end{equation}}
\def\noi{\noindent}

\begin{document}
\maketitle
\medskip
\noi{\bf Abstract.}
Calculations of the production rate of particles with $x>1$
in nuclear collisions due to the interaction of colour strings are
presented.  Momentum and colour sum rules are used to determine the
fragmentation functions  of fused strings. Mechanisms of the
string interaction are considered with total and partial overlapping
in the transverse plane. 
The results reveal strong dependence of the chosen 
mechanism. In the percolation scenario with partial overlapping  
the $x$-dependence of the production rates  agrees well with the existing 
data. The magnitude of the rates for $\pi^+$ production is in agreement
with experiment. However the rates for the protons are substantially below
the data.

\section{Introduction}
Production of particles in nuclear collisions 
in the kinematical region prohibited in the free nucleon kinematics
("cumulative particles")  has
long aroused interest both from the theoretical and pragmatic point
of views. On the pragmatic side, this phenomenon, in principle, allows to
raise the effective collision energy far beyond the nominal accelerator one.
This may turn out to be very important in the near future, when all
possibilities to construct still more poweful accelerator facilities
become exhausted. Of course one should have in mind that 
the production rate
falls very rapidly above the cumulative threshold, so that to use
the cumulative effect for practical purposes high enough luminosities are
necessary. On the theoretical side, the cumulative effect explores the
hadronic matter at high densities, when two or more nucleons overlap
in the nucleus. Such dense clusters may be thought to be in a state
which closely resembles a cold quark-gluon plasma. Thus cumulative
phenomena could serve as an alternative way to produce this new
state of matter.

There  has never been a shortage of models to describe the cumulative
phenomena, from the multiple nucleon scattering mechanism to
repeated hard interquark interactions [1]. However it should be
acknowledged from the start that the cumulative particle production
is at least in part a soft phenomenon. So it is natural to study it within
the models which explain successfully soft hadronic and nuclear
interactions in the non-cumulative region. Then one could have a
universal description of particle production in all kinematical
regions. The non-cumulative particle production is best described by
the colour string models, in which it is assumed that during the
collisions colour strings are stretched between the partons of colliding
hadrons (or nuclei), which then decay into more strings and finally
into observed produced hadrons

It is clear that if the  strings are formed
independently for different partons, all the cumulative effect can
be generated exclusively by the partonic movement. This can lead to 
cumulative particles due to 
the Fermi-motion of the nucleons in the participant nuclei. 
However in all realistic nuclear models the Fermi-motion generates
cumulative particles only in the region immediately above the
cumulative threshold. Beyond the threshold the effect dies so quickly
that there is absolutely no hope to explain by the Fermi-motion
the experimentally known production rates at $x\sim 2\div 3$ .
Thus in the string models the cumulative effect is related to the
interaction between strings, in particular, to their fusion, which
creates strings of higher energy and thus of a longer length in rapidity.
A model of string fusion and their percolation was proposed by the authors 
some
time ago [2]. It proved to be rather successful in explaining a series of
phenomena related to collective effects among the produced strings, such as
damping of the total multiplicity and strange baryon enhancement.
Calculations of the production rates
in the cumulative region made by the Monte-Carlo algorithm allowing for
fusion of only two strings gave encouraging results [3]. They agree 
quite well with the existing data  for hA collisions at 
$E_{cm}=27.5$ GeV [4,5 ]. However to
pass to higher energies and heavy-ion collisons one has to consider
a possibility of interaction of many strings.
In this note we study such an interaction using a simplified model
in which both colour and energy-momentum conservation are
imposed on the average.

From the start it is not at all obvious that the colour string
approach may give reasonable results in the deep fragmentation region,
near the kinematical threshold. The string picture has been introduced mainly
to describe particle production in the central region, where its results 
 agree with the data very well. It turned out however that it gives 
a quite reasonable description of the pion production also in the
fragmentation region. The baryon spectra
in the nucleus fragmentation region, on the contrary, are very poorly 
described, which circumstance is standardly ascribed  to the interaction 
between nucleons as a whole [6]. So, moving still 
further into a cumulative region, we may only expect a reasonable description
for the  production of pions, and not of  
baryons. As we shall see from our results, this is indeed so. 
We  get a very reasonable description 
of the pion production rates  for  $1<x<2$ at 27.5 GeV [5] but we are not 
able to describe the proton rates, which are experimentally
two orders of magnitude greater than our predictions. Obviously the bulk 
of cumulative protons come from a different mechanism, which does not involve
colour string formation but rather interactions of nucleons as a whole.
Such a mechanism was included in the Monte-Carlo code of [3], which is
the reason why it gave results for nucleon production in agreement
with the experimental data.

\section{The model}
The fully extended string model assumes that each of the colliding
hadrons consists of partons (valence and sea quarks), distributed both
in rapidity and transverse space with a certain probability, deduced
from the experimentally known transverse structure and certain theoretical
information as to the behaviour of the $x$ distributions at $x=0$ and
$x=1$. These distributions are taken to be the ones for the endpoints
of the generated strings. As a result, the strings aquire a certain length
in rapidity. We shall choose the c.m. system for the colliding hadrons
with the nucleus (projectile) moving in the forward direction. The cumulative
particles thus will be observed in the forward hemisphere.

Let a parton from the projectile carry a part
$x_{1+}$ of the "+" component of its momentum $p_1$ and
a partner parton from the target carry a part $x_{2-}$ of the "-" component of
its momentum $p_2$. The total energy squared for the colliding
pair of nucleons is 
\beq S=2p_{1+}p_{2-}=m^2e^Y\eeq
where $m$ is the nucleon mass and $Y$ is the total rapidity available.
The c.m. energy squared  accumulated in the string is then
\beq s= x_{1+} x_{2-}S\eeq
Note that the concept of a string has only sense in the case when $s$
 is not too small, say more than $m^2$. So both $ x_{1+}$ and $x_{2-}$ cannot be
too small.
\beq  x_{1+}, x_{2-}>x_{min}=m/\sqrt{S}=e^{-Y/2}\eeq
Correspondingly we relate the scaling variables for the string endpoints 
to their rapidities by
\beq
y_1=Y/2+\ln x_{1+},\ \ y_2=-Y/2-\ln x_{2-}\eeq
Due to (3) $y_1\geq 0$ and $y_2\leq 0$. The "length" of the string is just
the difference $y_1-y_2$. 

Standardly it is assumed that the spectrum of
observed particles generated by the string is nearly a constant along its length
and zero outside. Due to the partonic distribution in $x$ the strings have
different lengths and moreover can take different position respective to
the center $y=0$.
The sea distribution in a hadron is much softer than the valence one.
In fact the sea distribution behaves as $1/x$ near $x=0$, so that the
average value of $x$ for sea partons is of the order $1/Y$ (see Eq. (3)).
As a result, strings attached to sea partons in the projectile nucleus
carry very small parts of longitudinal momentum in the forward direction, 
which
moreover fall with energy, so that they seem to be useless for building 
up the cumulative particles. 
This allows us to  retain only strings
attached to valence partons, quarks and diquarks, in the projectile and 
neglect all  strings attached to sea quarks altogether.
Note that the number of the former is exactly equal to $2A$ and does not
change with energy. So for a given nucleus we shall have a fixed number of
strings, independent of the energy. 

The upper end rapidities of the 
strings attached to diquarks are usually thought to
be  larger than of those attached to the quarks, since the average
value of $x$ for the diquark is substantially larger  that for the
quark. Theoretical considerations lead to the conclusion  that as 
$x\to 1$ the distributions for the quark and diquark in the nucleon
behave as $(1-x)^{3/2}$  and $(1-x)^{-1/2}$ respectively, modulo
logarithms [6]. Neglecting the logarithms and taking also in account the
behaviour at $x\to 0$ we 
assume that these distributions are
\beq
q(x)=\frac{8}{3\pi}x^{-1/2}(1-x)^{3/2}
\eeq
for the quark and
\beq
qq(x)=q(1-x)=\frac{8}{3\pi}x^{3/2}(1-x)^{-1/2}
\eeq

The quark and diquark strings will be attached  to all sorts of partons 
in the target nucleon: valence
quark and diquark and sea quarks. Their position in rapidity in the backward
hemisphere will be very different.  However we are
 not interested in the spectrum in the backward hemisphere. So, for our 
purpose, limiting ourselves with the forward hemisphere, we 
may take lower  ends of the strings all equal to $x_{min}<<1$.
As a result, in our  model at the start we have $N$ initially created
strings, half of them attached to quarks and  half to diquarks, their 
lower ends in rapidity all equal to \[y_2=Y/2+\ln x_{min}\]
and their upper ends distributed in accordance with (5) and (6).
As soon as they overlap in the transverse space they fuse into new
strings with more color and more energy. This process will be studied in
the next section.    

\section{String fusion and conservation laws}
\subsection{Complete fusion}
Let $n$ strings overlap completely in the transverse area and form a new 
string of higher colour. The process of fusion obeys two conservation
laws: those of colour and momentum.
As a result of the conservation of colour, the colour of the fused string is
$\sqrt{n}$ higher than that of the ordinary string [7,8]. From the 4 momentum
conservation laws we shall be interested mostly in the conservation of the
"+" component, which leads to the conservation of $x$.  The fused string 
will have the upper
endpoint with $x_n=\sum_{i=1}^nx^{(i)}$, where $x^{(i)}$ are upper ends of 
fusing strings(we omit the subscripts "1+", 
since we shall be interested only in these variables in the future).

These properties of the fused string transform into certain sum rules which
restrict a possible form of the spectrum of produced hadrons. Let us first
assume that only one sort of particles is produced and
let the mutiplicity density ("fragmentation function" in the
terminology of [6])  of the fused 
string be \[ \tau_n=\frac{d\mu_n}{dy}\]
where $\mu$ is the multiplicity. As mentioned, the total
number of particles produced in the forward hemisphere by the fused string 
should be $\sqrt{n}$ greater than by the ordinary string. This leads to
the multiplicity sum rule:
\beq
\int_{x_{min}}^{x_n}\frac{dx}{x}\tau_n(x)=\frac{1}{2}\mu_0\sqrt{n}
\eeq
where we denote $\mu_0$ the total multiplicity from a simple string.
The produced particles have to carry all the longitudinal momentum in the
forward direction. This results  in the  sum rule for $x$:
\beq
\int_{x_{min}}^{x_n}dx\tau_n(x)=x_n
\eeq
In these sum rules $x_{min}$ is given by (3) and is small.
Passing to the scaled variable $z=x/x_n$ we rewrite the two sum rules
as
\beq
\int_{z_n}^1\frac{dz}{z}\tau_n(z)=\frac{1}{2}\mu_0\sqrt{n}
\eeq
and
\beq
\int_{z_n}^1dz\tau_n(z)=1
\eeq
where
\beq
z_n=x_{min}/x_n
\eeq

These sum rules put severe restrictions on the form of the distribution
$\tau_n$, which obviously cannot be independent of $n$. Comparing (7) and
(8) we see that the spectrum of the fused string has to vanish at its
upper threshold faster than for the
simple string. In the scaled variable $z$ it is shifted to smaller values 
(and thus to the central region).
This must have a negative effect on the formation of cumulative particles
produced at the extreme values of $x$.

To proceed, we choose a simplest form for the distribution $\tau_n$:
\beq
\tau_n(z)=a_n(1-z)^{\alpha_n-1}
\eeq
The $x$ sum rule relates $a_n$ and $\alpha_n$:
\beq
a_n=\alpha_n(1-z_n)^{-\alpha_n}\simeq \alpha_n
\eeq
The multiplicity sum rule finally determines $\alpha_n$:
\beq
\alpha_n(1-z_n)^{-\alpha_n}\int_{z_n}^1\frac{dz}{z}(1-z)^{\alpha_n-1}=
\frac{1}{2}\mu_0\sqrt{n}
\eeq

These equation can be easlily solved when $z_n\to 0$
We present the integral in (14) as
\beq
\int_{z_n}^1\frac{dz}{z}[(1-z)^{\alpha_n-1}-1]+\ln\frac{1}{z_n}
\eeq
The integral term is finite at $z_n=0$ so that we can write it as a 
difference
of integrals in the intervals $[0,1]$ and $[0,z_n]$. The first can be found
exactly 
\[
I_1=\int_0^1\frac{dz}{z}[(1-z)^{\alpha_n-1}-1]=
\lim_{\epsilon\to 0}
\int_0^1dzz^{-1+\epsilon}[(1-z)^{\alpha_n-1}-1]=\]\beq
\lim_{\epsilon\to 0}\Big[{\rm B}(\alpha_n,\epsilon)-\frac{1}{\epsilon}\Big]=
\psi(1)-\psi(\alpha_n)
\eeq
The second term has an order
$-(\alpha_n-1)z_n$
and is small unless $\alpha_n$ grows faster than $n$, which is not the case
as we shall presently see. In fact we shall find that $\alpha_n$ grows roughly
as $\sqrt{n}$, which allows to neglect the second factor in (14) and rewrite
it in its final form
\beq
\alpha_n \Big[\ln\frac{1}{z_n}+\psi(1)-\psi(\alpha_n)\Big]=
\frac{1}{2}\mu_0\sqrt{n}
\eeq
Obviously $\alpha_n$ grows as $\sqrt{n}$, modulo logarthmic
dependence, as  mentioned. To finally fix the distributions we have to
choose the value of
$\alpha$ for the simple string. We take the simplest choice $\alpha_1=1$
for an average string with $x=x_0=1/2$,
which corresponds to a completely flat spectrum and agrees with the 
results of [6].  This fixes the multiplicity density for the average string
\beq\tau_1(y)=1\eeq
which favorably compares to the value 1.1 extracted from the experimental data
[8]. 

With (18) we find from (17)
\beq
\frac{1}{2}\mu_0=\ln\frac{1}{z_0},\ \ z_0=\frac{x_{min}}{x_0}
\eeq
and the equation (17) can be rewritten as
\beq
\alpha_n \Big[\ln\frac{n}{z_0}+\psi(1)-\psi(\alpha_n)\Big]=
\sqrt{n}\ln\frac{1}{z_0}
\eeq

One concludes that at $S\to\infty$ the solution is $\alpha_n=\sqrt{n}$.
However this is modified by logarithmic terms when $n$ is high enough:
$n\sim \sqrt{S}/m$. At finite $S$ Eq. (20) can be solved numerically for
$\alpha_n$

In any case, we find that with the growing $n$ the spectrum of produced
particles
goes to zero at $z\to 1$ more and more rapidly. So although strings with large
$n$ produce particles with large values of $x\le x_n$, the production rate
is increasingly small.

\subsection{Partial fusion}
Now we pass to a more difficult case when two or more strings only overlap
partially. This case corresponds to weaker interaction between strings,
which retain their form in the transverse space. This case
 lies at the basis of the percolation phenomenon.
To start, let us note that for two or several completely independent strings
the conservation laws and the following sum rules are automatically
satisfied if they are satisfied for each string individually.
As a consequence the case of partially fusing strings has to be treated 
differently depending on whether we consider a formed cluster as a single
string or as a set of many strings formed by the various areas where
a fixed number of particular strings overlap. In the first case we have
to impose a single pair of conservation laws for the whole cluster. In the
second case we find many pairs of conservation laws for strings
formed by different overlaps.

An  approach which admits the most direct physical interpretation
is to consider
a cluster of
$n_c$ strings as a set of many independent ``ministrings'' formed by different
overlaps. This case corresponds to the minimal interaction between strings,
which actually only superimpose in the transverse space without changing
any of their properties. 
In this case the conservation laws and sum rules have to be imposed
for each particular overlap. If the area of a given overlap of $n$ strings
is $S_n^{(i)}$, where $i$ enumerates different overlaps of $n$ strings,
then the colour of the corresponding ministring is [8]:
\beq
Q_n^{(i)}=\sqrt{n}\frac{S_n^{(i)}}{C_1}Q_0
\eeq
where $Q_0$ is the colour if the simple string and $C_1$ is its area.. As a 
result, the total multiplicity will be changed correspondingly.
Passing to the momentum, we have to assume some manner in which the total
$x=x_c$ of the cluster is distributed among various ministrings formed by 
overlaps.
A natural way (similar to the one used for the distributiuon of color)
is to assume that the part of the longitudinal momentum shared by a string
in a particular overlap is proportional to the area of the latter. Then
the total $x$ of the overlap is:
\beq
x_n^{(i)}=x_c\frac{nS_n^{(i)}}{n_cC_1}
\eeq
where  $n_c$ is the number of 
strings forming the cluster. Indeed, since
\beq
\sum_{i,n}nS_n^{(i)}=n_cC_1
\eeq
we have for each cluster
\beq
\sum_{i,n}x_n^{(i)}=x_c
\eeq
as it should be. Introducing for each individual overlap its scaled variable
$
z=x/x_n^{(i)}
$ we shall write the momentum sum rule in the same form (10) as before.   
The multiplicity sum rule will now
read
\beq
\int_{z_n^{(i)}}^1\frac{dz}{z}\tau_n(z)=
\frac{1}{2}\mu_0\sqrt{n}
\frac{S_n^{(i)}}{C_1}
\eeq
where
\[ z_n^{(i)}=x_{min}/x_n^{(i)}.\]
With the choice (12) the final equation for $\alpha_n$ will take the form
\beq
\alpha_n \Big[\ln\frac{1}{z_n^{(i)}}+\psi(1)-\psi(\alpha_n)\Big]=
\ln\frac{1}{z_0}\sqrt{n}\frac{S_n^{(i)}}{C_1}
\eeq
Obviously calculations
cannot be done analytically now but require numerical simulation.
They result very time consuming, since one has to identify
each particular overlap in each cluster.

A simpler alternative is to treat the whole cluster  as a single
string of a complicated form and  area $C_n$,
 with properties intermediate between
$n$ separate strings and completely fused ones. This implies a 
considerable amount of interaction between strings, which redistributes the 
colour 
and momentum homogeneously over the cluster area. Let the cluster  of 
$n$ strings have the total
$x$ equal to $x_c$. However the maximal momentum $x_n$ of the emitted 
particle cannot be equal to
$x_c$, since for $n$ separate strings just touching each other it is only 
$x_1$. So it should be intermediate between $x_1$ and $x_c$, depending on
the fusion intensity. A natural choice seems to be
\beq x_{n}=x_c\frac{C_1}{C_n}\eeq
In fact for $n$ independent strings $C_n=nC_1$ and $x_{n}=x_c/n$,
which is just an average of the maximal momenta of fusing strings.
For $n$ completely fused strings $C_n=C_1$ and $x_n=x_c$ as it should be.
The momentum sum rule is then
\beq
\int_{x_{min}}^{x_n}dx\tau_n(x)=x_c
\eeq
Taking 
\beq
z=\frac{x}{x_n},\ \ \tau=\frac{x_c}{x_n}\tilde{\tau}
\eeq
we obtain an equation for $\tilde{\tau}$ of  the same form as (25) with
$
z_n=x_{min}/x_n
$
The
total multiplicity will be also intermediate between the multiplicity of $n$
independent strings and of the completely fused string. A reasonable choice 
similar to (27) is [9]
\beq
\mu_c=\mu_0\sqrt{n\frac{C_n}{C_1}}
\eeq
This will lead to the corresponding change in the multiplicity sum rule,
which in terms of $z$ and $\tilde{\tau}$  becomes
\beq
\int_{z_n}^1\frac{dz}{z}\tilde{\tau}_n(z)=
\frac{1}{2}\mu_0\sqrt{n\frac{C_1}{C_n}}
\eeq
For  $n$ independent strings  (31) goes into (7) 
with $n=1$.  For complete
overlapping, when
$C_n=C_1$,  (30) reproduces (7). With the choice (12) for the distribution in
rapidity we now get an equation for $\alpha_C$
\beq
\alpha_C \Big[\ln\frac{1}{z_n}+\psi(1)-\psi(\alpha_C)\Big]=
\ln\frac{1}{z_0}\sqrt{n\frac{C_1}{C_n}}
\eeq
This equation has to be solved separately for each cluster, so that the form of
the 
distribution of particles produced by a cluster of $n$ strings will depend on
its geometry (in fact only on its total area). The numerical calculations 
look easier now, since one  has to identify individual clusters only.

\subsection{Various types of hadrons}
In reality various types of hadrons are produced. In the cumulative 
region the relevant particles are nucleons and pions, the production
rates of the rest being much smaller. 
The multiplicity densities for each sort of hadrons will obviously
depend on the flavour contents of the fused strings, that is, of the
number of quark and diquark strings in it. 

Consider the case of complete overlapping of $n$ strings. Let the string be 
composed of $k$ quarks and $n-k$ diquarks.
 We shall then have distributions 
$\tau_{nk}^{(h)}$ for the produced hadrons $h$.
Obviously the multplicity and momentum sum rules are now insufficient to
determine each of the distribution $\tau_{nk}^{(h)}$ separately.
A possibility to overcome this difficulty may consist in using the
Regge phenomenology or quark power counting rules to determine the behaviour 
of the distribution near its kinematical threshold $x\rightarrow x_n$.
However for multiquark configurations this  seems  too complicated and
insecure, since this behaviour is then governed by Regge cuts rather than 
poles.
Therefore we shall adopt a simpler approach. In our picture the observed
hadron is produced when the  parton (quark or diquark) emerging from 
string decay neutralizes  its colour by picking up an appropriate parton
from the vacuum. In this way a quark may go into a pion if it picks up
an antiquark or into a nucleon if it picks up two quarks. The quark 
counting rules 
tell  us that the behaviour at the threshold in the second case will
have two extra powers of $(x-x_n)$. Likewise a diquark may go either
into a nucleon picking up a third quark or into two pions picking up
two antiquarks, with a probability smaller by a factor $(x-x_n)^2$ at
the threshold. Since it is the threshold behaviour which is responsible for
the cumulative particles, we shall assume, as a first approximation, that
the quarks go exclusively into pions and diquarks go exclusively into
nucleons. As a result, the ratio of produced pions to nucleons will
be equal to that of quarks to diquarks, that is, $k/(n-k)$.

This means that we may proceed as before assuming  one sort of hadrons 
and in the end just multiply the obtained distributions $\tau_{nk}$ by
the appropriate factors
\beq
\tau_{nk}^{(h)}(x)=\xi_{nk}^{(h)}\tau_{mk}(x),
\eeq 
where for the pions and nucleons
\beq
\xi_{nk}^{(\pi)}=\frac{k}{n},\ \ \xi_{nk}^{(N)}=\frac{n-k}{n}
\eeq
Obviously this implies that we have to track the flavour contents of the 
fused strings in the Monte-Carlo simulations.

\section{Numerical results}
We shall study different scenarios for the interaction between colour 
strings. 
We start with the scenario of string fusion,
when as soon as the 
strings touch each other they fuse into a new one with a higher colour 
and of the same form and transverse area as the initial ones.
This corresponds to the maximal 
interaction between
strings, which change not only their colour and momentum but also its 
geometrical form.
This
scenario was proposed  and realized in a fully developed 
Monte-Carlo algorithm in [2]. Its application to particle production has 
given very satisfactory results for energies ranging from SPS to RHIC [10].

In this case we do not have to bother about partial overlapping, so 
that Eq. (20) can be used directly to determine the multiplicity of a string
with an arbitrary high colour. The distribution of the strings in colour 
can be deduced analitically for high enough string densities in this
scenario [11]. It is governed by a dimensionless parameter 
\beq
\eta=\frac{NC_1}{C}
\eeq
where $C$ is the total transverse area of the interaction and $N$ the total
number of strings. 
For a 
given $\eta$ and large $N$ one finds that the strings are distributed in 
colour according to the Poisson law:
\beq
\langle\nu_n\rangle=Ne^{-\eta}\eta^{n-1}/n!
\eeq
So with string fusion we find the total multiplicity  
 as
 \beq
\frac{d\mu}{dy}=N\sum_n\frac{1}{n!}e^{-\eta}\eta^{n-1}\alpha_n
\left(1-\frac{x}{x_n}\right)^{\alpha_n-1},
\eeq
where the sum goes over all colors $n$ such that $x<x_n$.
Parameters $\alpha_n$ are to be calculated from Eq. (20)
and the string momentum $x_n$ and its flavour composition
(number of quarks and diquarks) has to be found 
 by Monte-Carlo simulations using  the distributions (5) and (6). 

The multiplicity results very weakly 
dependent on the energy through the  value of $x_{min}$ (eq. (3)),
this dependence practically absent for energies above 200 GeV.
Our results for $d\mu/dy$ per string at the LHC energy 6 TeV (Y=17.5) for 
different values of $\eta$ are presented in Fig. 1 and 2 for pions and 
nucleons respectively.

\begin{figure}
\centerline{ \epsffile{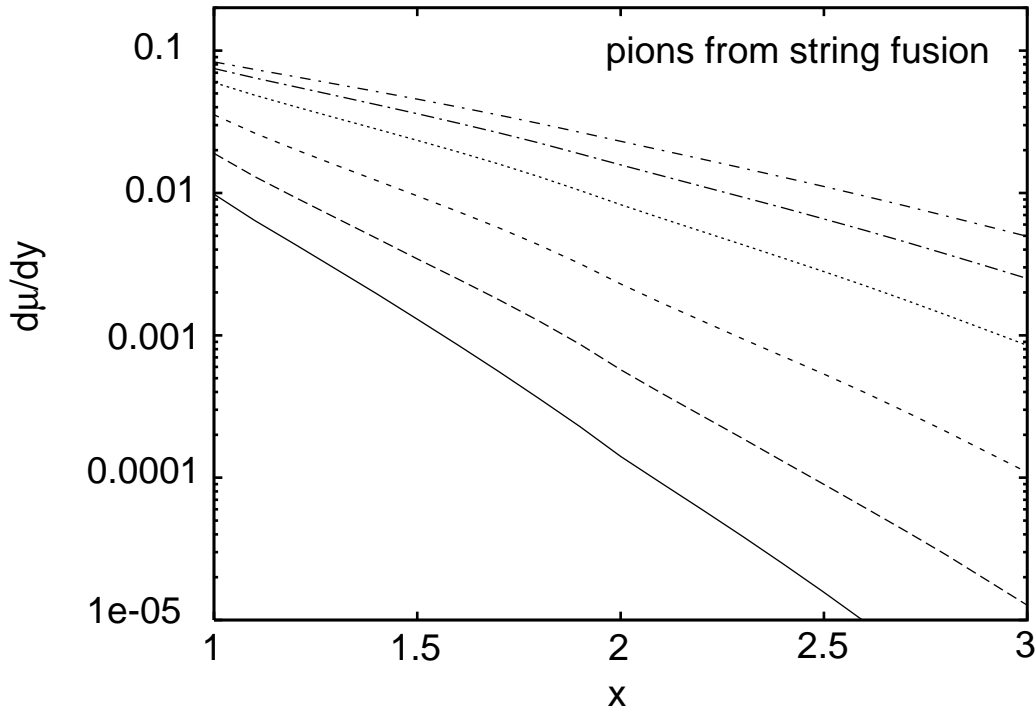}}
\caption{Multiplicity density $d\mu/dy$ per string for pions in the 
fusion scenario for $Y=17.5$.
The curves from bottom upwards correspond to $\eta=0.25$, 0.5, 1, 2, 3 
and 4.}
\end{figure}

\begin{figure}
\centerline{ \epsffile{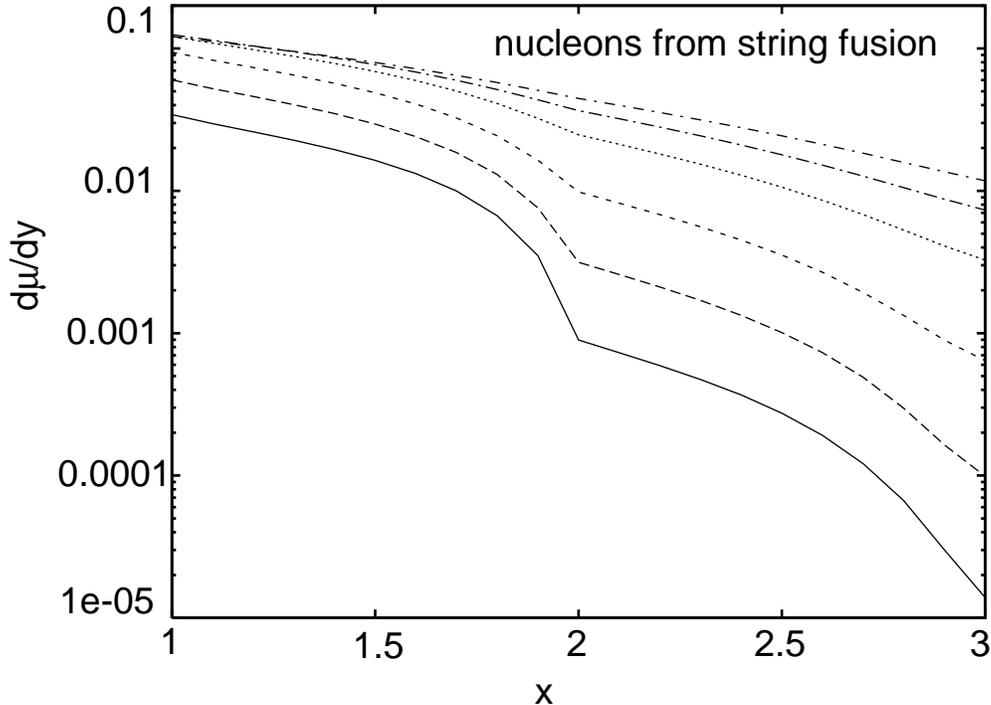}}
\caption{Same as Fig. 1 for nucleons.}
\end{figure} 

In an alternative scenario, which lies at the basis of the percolation 
phenomenon ("percolation scenario") strings are allowed to overlap partially
and form clusters of different forms and number of simple strings. 
The total multiplicity will then depend on the chosen model of the 
distribution of colour and momentum within a given cluster, as explained in
the previous section.

If a cluster is assumed to be split in various  ministrings formed by
different separated overlaps, the total multiplicity will be a sum of
contributions from all ministrings
\beq
\frac{d\mu}{dy}(x)=\sum_{n,i}\tau_{n,i}\left(\frac{x}{x_n^{(i)}}\right)
\eeq
where for $z<1$  $\tau_{n,i}(z)$ is given by (12) with parameters 
$\alpha_{n,i}$ determined from Eq. (26) and
 for $z>1$ $\tau_{n,i}(z)=0$. 
If a cluster acts as a single string with an averaged distribution 
$\tau_C(z)$ described in a previous section, then one has the same
expression (38), where the sum is now extended over all clusters.
The parameters $\alpha_C$ are now to be determined from Eq. (32).
In both cases analytical calculations are not possible and one has to 
recur to Monte Carlo simulations. They simulate the 
geometrical distribution of 
strings in the interaction area and their $x$ and flavour contents.
Afterwards one has to identify all overlaps of a given number of strings
or clusters of strings and their areas.
The results (per string) for the two possibilities are presented in 
Figs. 3,4 and 5,6 respectively.

\begin{figure}
\centerline{ \epsffile{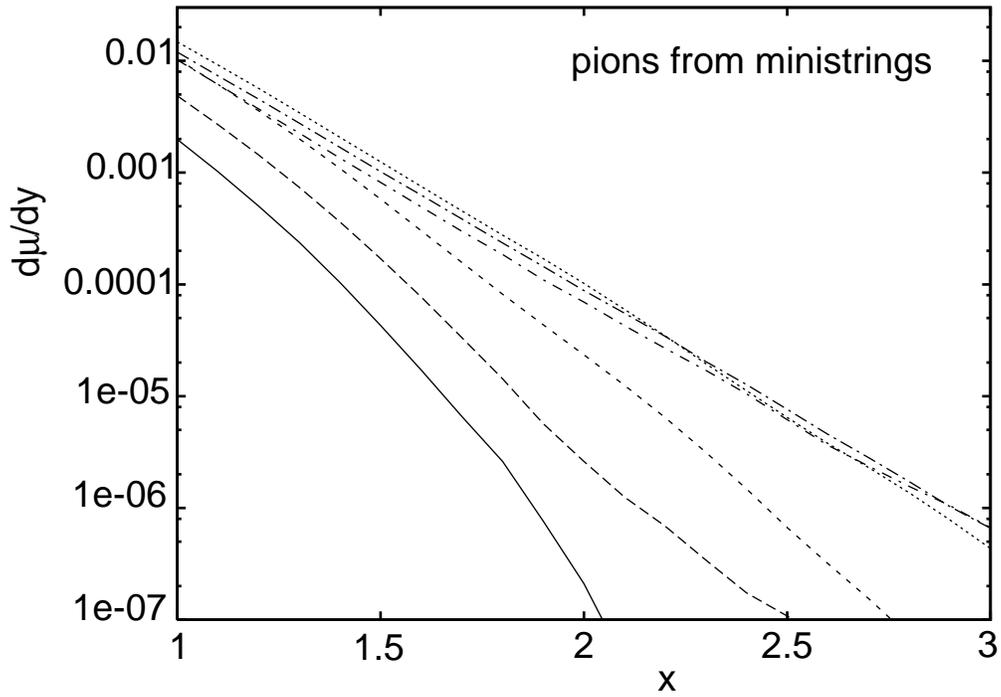}}
\caption{Multiplicity density $d\mu/dy$ per string for pions in the 
percolation scenario with ministrings (overlaps) as emitters 
at $Y=17.5$. At $x=2$ the curves from bottom upwards correspond to
$\eta=0.25$, 0.5, 1, 4, 3 and 2.}
\end{figure}

\begin{figure}
\centerline{ \epsffile{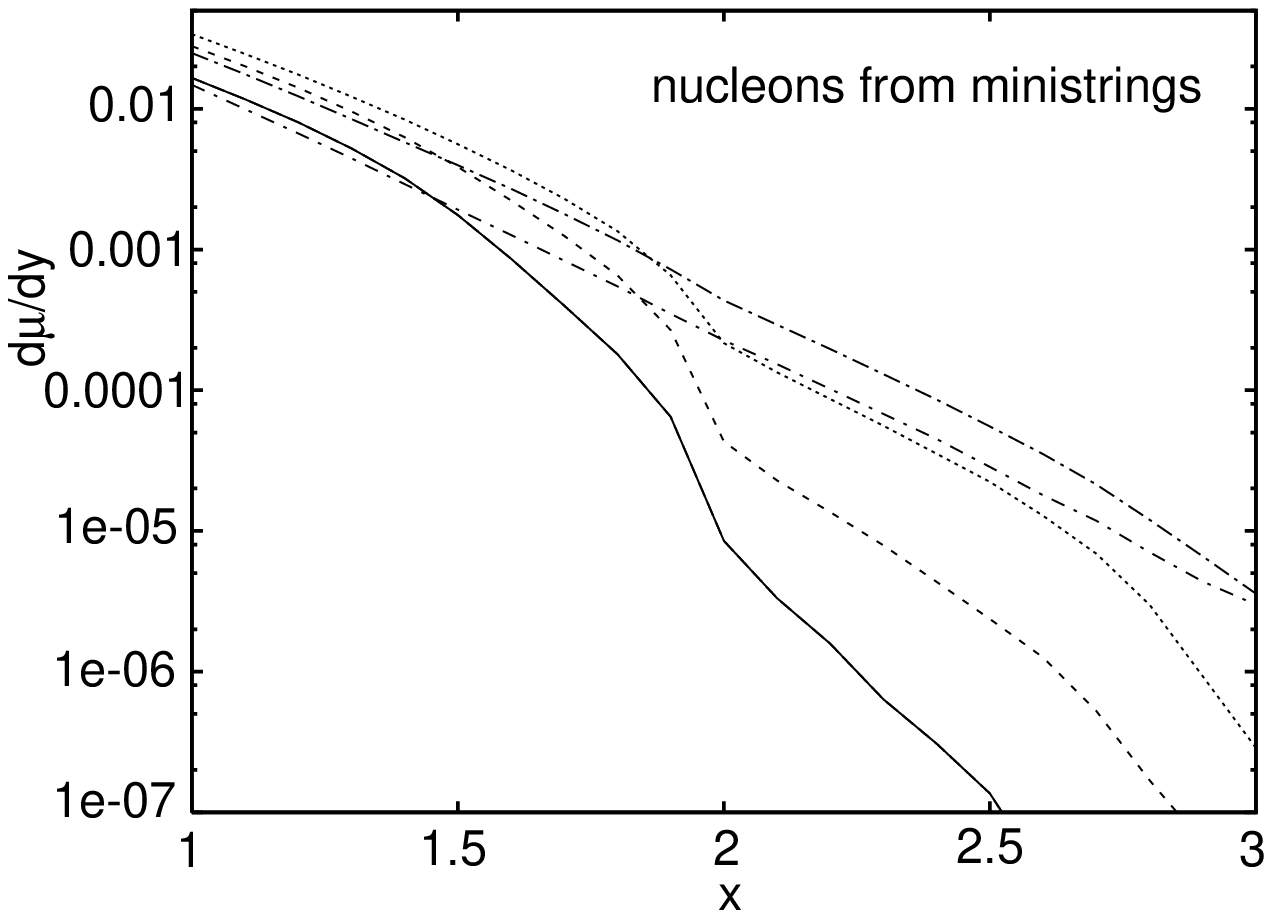}}
\caption{Same as Fig. 3 for nucleons. At $x=2.5$ the curves from bottom
upwards correspond to $\eta=0.25$, 0.5, 4, 1, 3 and 2.}
\end{figure}

\begin{figure}
\centerline{ \epsffile{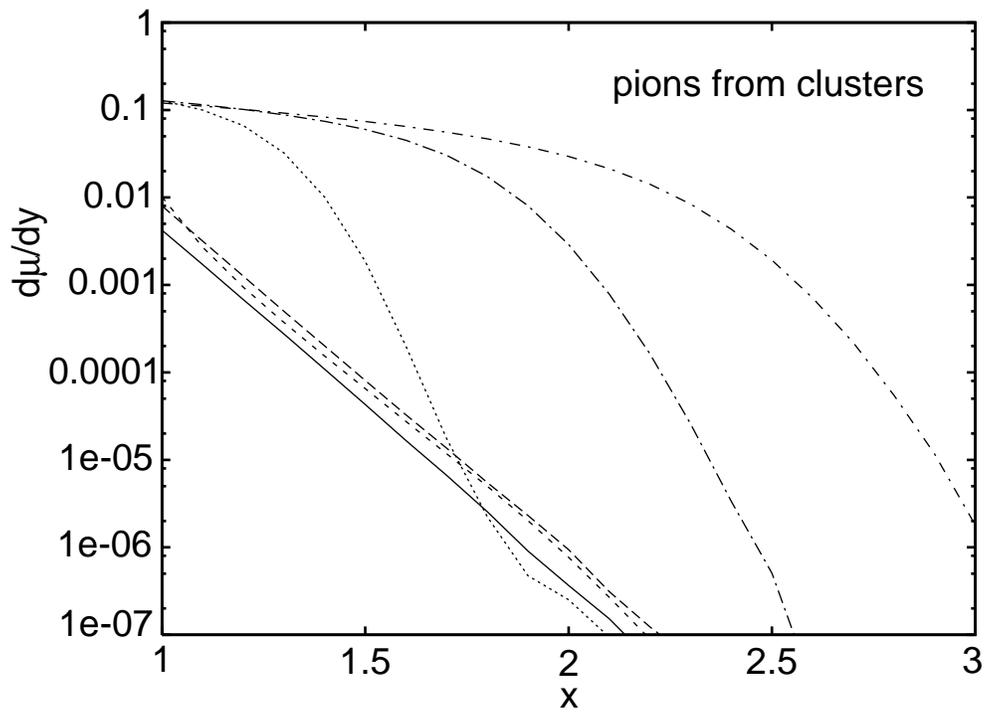}}
\caption{Multiplicity density $d\mu/dy$ per string for pions in the 
percolation scenario with clusters as emitters at 
$Y=17.5$.  At $x=2$ the curves from bottom upwards correspond to
$\eta=2$, 0.25, 1, 0.5, 3 and 4.}
\end{figure}

\begin{figure}
\centerline{ \epsffile{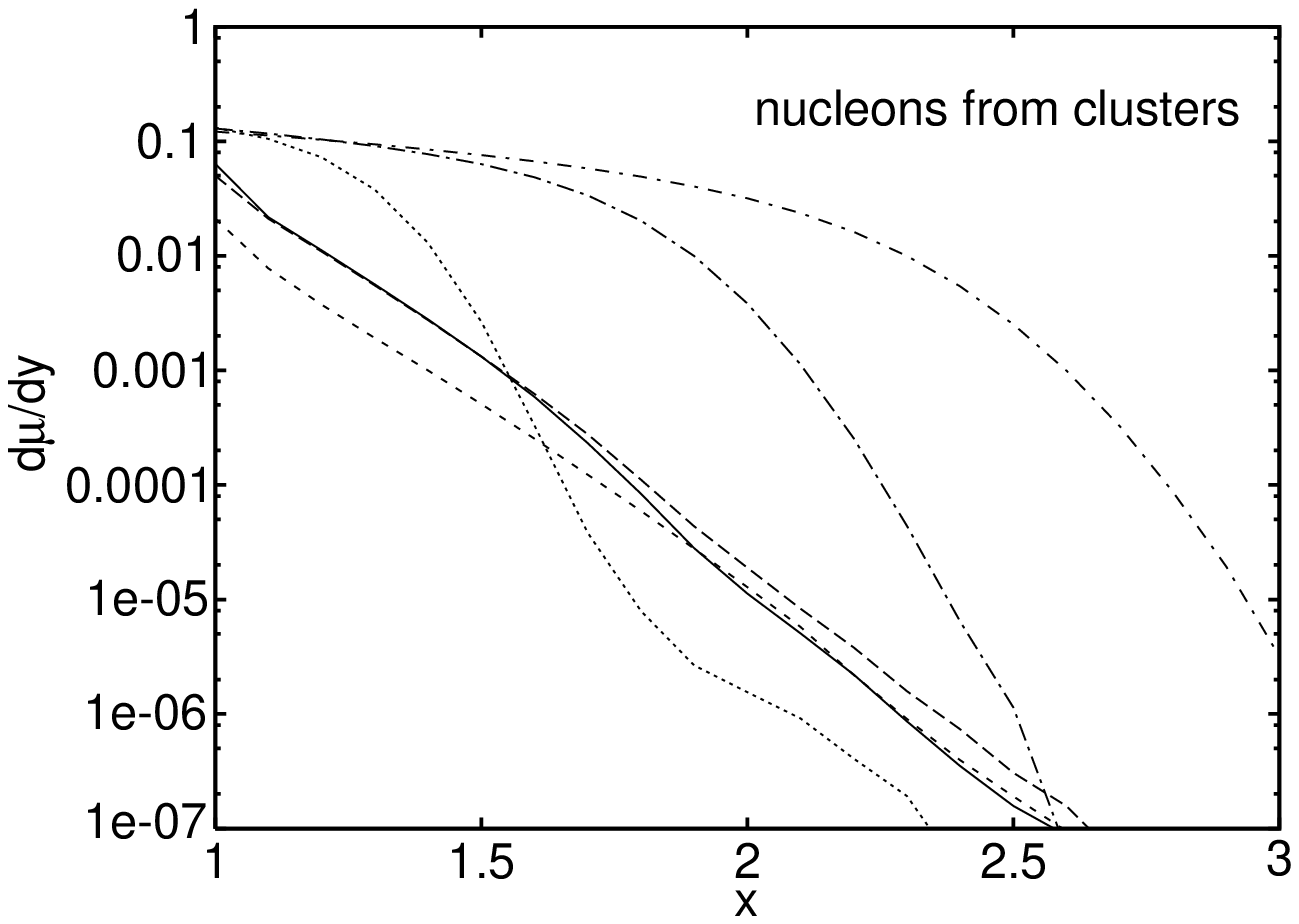}}
\caption{Same as Fig.5 for nucleons.  
At $x=2$ the curves from bottom upwards correspond to
$\eta=2$, 0.25, 1, 0.5, 3 and 4.}
\end{figure}

One observes that probability of cumulative production  strongly 
depends on a chosen scenario. 
The fusion scenario leads to higher 
cumulative particle production rates, which also fall with $x$ much slowlier
than in the percolation scenario. 
The $x$-dependence of the rates for pion production is rather well 
described by the standard parametrization 
 \beq
\frac{d\mu}{dy}=Ce^{-\alpha x}
\eeq
where the slope $\alpha$ steadily falls with $\eta$ from
$\alpha\simeq 4$ at $\eta=0.25$ down to $\alpha\simeq 1.5$ at $\eta=4$.
At fixed $x$ the rates steadily grow with 
$\eta$. For the nucleon production and low values of $\eta\le 1$ the 
parametrization (39) holds only for the regions $1<x<1.7$ and
$2<x<2.5$ where   $\alpha\sim 2$. In between and at $x>2.5$ the 
slopes rise up to $4\div 6$ indicating an abrupt change of $x$-behaviour
at $x\simeq 2$ which gradually disappears with the growth of $\eta$.
At $\eta\geq 2$ the $x$-dependence is roughly described by (39) with the same
slope $\sim 1.5$ as for pions. 

In the percolation scenario at relatively small values of $\eta<1$
the results are qualitatively independent of the choice of emitters.
With both clusters and ministrings one finds the rates which fall with 
$x$ much faster than in string fusion. Again for pions the 
parametrization (39) holds, with
$\alpha$ slowly falling with $\eta$ from $\sim 9$ at $\eta=0.25$ to $\sim 
6.5$ at $\eta=1.$ for ministrings as emitters and from $ \sim 9.5$ at 
$\eta=0.25$ to $\sim 8.5$ at $\eta=1$ for clusters. For nucleon production
from ministrings a change of behaviour in the vicinity of $x=2$ is
again observed. With clusters this change is hardly perceptible. Outside
this region the slopes are somewhat smaller than for pions: $4\div 7$ for
ministrings and $7\div 9$ for clusters. 
 
At higher values of $\eta$, beyond the
percolation threshold $\eta_c\sim 1.1\div 1.3$,  the behaviour of the 
production rate begins to depend
crucially on the choice of emitters. With clusters as emitters
a radical change is observed and the production rate becomes 
close to the fusion scenario up to a certain 
value of $x$ when the rate abruptly goes to zero. The critical value of 
$x$ grows with $\eta $ and, as seen from Fig.3,   lies around 
1.5 at 
$\eta=2$ and at around 2.5 at $\eta=4$. On the other hand, the ministrings 
scenario does not show such an abrupt change of behaviour, the rates 
well described by (39) with a slope $ 4\sim 5$ both for pions and nucleons.  

To pass to physical inclusive cross-sections one has to know the number
of strings $N$ in a particular reaction. This  number is 
obtained by multiplying their number in the nucleon (two) by the number of
collisions $\nu$. 

In hA collisions in the nuclear fragmentation region 
one has
\beq
\nu=\frac{A\sigma_{pp}}{\sigma_{pA}}
\eeq
The interaction area is of the order 
$\sigma_{pp}$, so that (35) gives  
\beq 
\eta=2\frac{AC_1}{\sigma_{pA}} 
\eeq
According to our previous results [2,3,8] $C_1=1.26$ mb.
To find the inclusive cross-section one has to multiply $d\mu/dy$ per 
string by $N$ and $\sigma_{pA}$ and divide by the number of
isospin components,
 which gives the inclusive cross-section per nucleon for positive pions and 
protons 
\beq
\frac{1}{A}\frac{d\sigma_{pA}^{(\pi^+)}}{dy}
=\frac{2}{3}\sigma_{pp}\frac{d\mu^{(\pi)}}{dy},\ \ 
\frac{1}{A}\frac{d\sigma_{pA}^{(p)}}{dy}
=\sigma_{pp}\frac{d\mu^{(N)}}{dy}
\eeq

In AB interactions the number of collisions depends on the centrality.
In minimum bias collisions, similar to (40),
\beq
\nu=\frac{AB\sigma_{pp}}{\sigma_{AB}} 
\eeq
The interaction area is the average overlap area. For $B<<A$ it is
roughly equal to $\sigma_{pB}$. For $A=B$ it is approximately equal
to $0.78 R_A^2$.  The inclusive cross-section per nucleon will be given 
by the same Eq. (42).
the only difference being that the value of $\eta$ should be calculated 
with (43) and so substancially higher than in $hA$ collisons. 
One can also easily deduce corresponding formulas for the cross-sections in
collisions with a given centrality, that is, at a given impact parameter $b$
(see [11]).

Passing to the comparison of our predictions with the experimental
data we have first to stress that all the existing  data refer to
the hadron nucleus collisions at very moderate c.m. energies  of 27.5 GeV 
and below. The strings which are created at these energies have mostly rather
small length in rapidity, so that effects of the energy division
between strings, neglected in our approach, become important. This implies
that our picture can  describe these experiments only rather crudely.

Still, forgetting for the moment the absolute magnitudes of the 
cross-sections, we conclude from Figs. 1-6 that the $x$-dependence 
is correctly reproduced in the two percolation scenarios, which
at small $\eta$ corresponding to these energies give the inclusive 
cross-section of the form (39) with the slope $\alpha\sim 6\div 7$ in 
complete agreement with the experimental values.
It is remarkable that this value of the slope is obtained practically
with no parameters, on a pure geometrical basis.
The string fusion scenario, in contrast, leads to much smaller values
of $\alpha$ which definitely contradict the experimental data.
So our first conclusion is that it looks as if the experimental
cumulative cross-section favour the percolation scenarios.

Passing to the absolute values, we note that in [4,5] 
the double differential cross-sections in momentum and angle are 
given. To convert them into $d\sigma_{pA}/dy$ we fitted the data with
\beq
\frac{1}{A}\frac{d\sigma_A(x,k_T^2)}{dyd^2k}=Ce^{-\alpha_{ex} x-\beta k_T^2}
\eeq
For positive pions produced on $Ta$ (A=181) at $\sqrt{s}=27.5$ and $x>1$
we get
\[ \alpha_{ex}=7.15,\ \beta=1.72\ \ {\rm and}\ \ 
C=124/ {\rm mb/ (GeV/c)}^2\]

The resulting "experimental" $d\sigma_A/Ady$ is a pure exponential in $x$ 
with slope $\alpha_{ex}$ and is shown as a straight line in Fig. 7. 
In the same figure we show our predictions for this case ($\eta=0.343$)
in the percolation scenarios choosing ministrings or clusters as emitters.
One observes that the agreement is quite reasonable, especially in view 
of the mentioned difficulties in applying our picture.

\begin{figure}
\centerline{ \epsffile{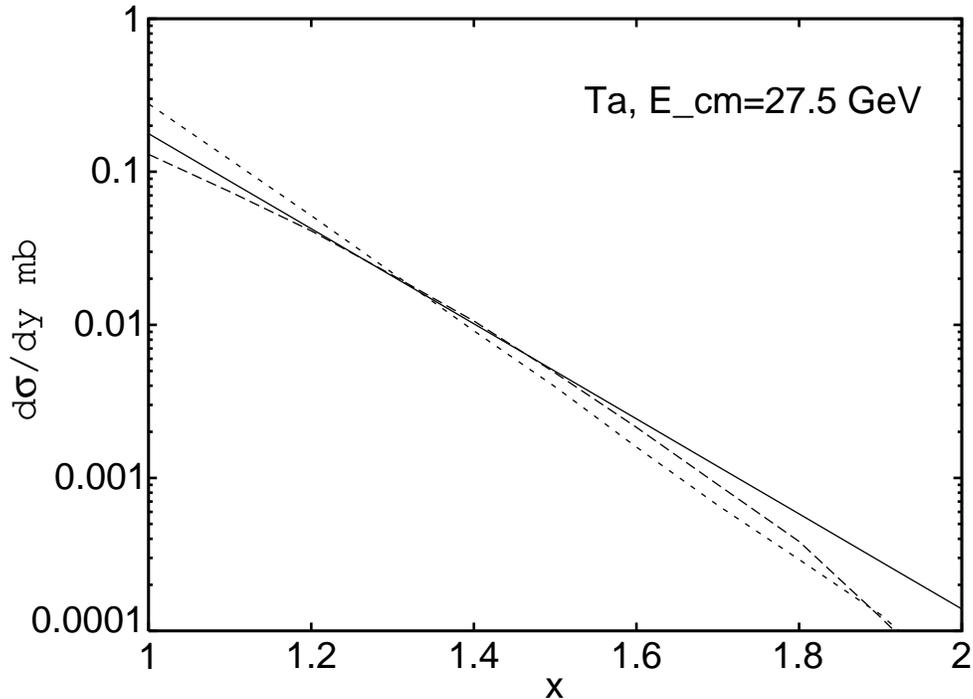}}
\caption{Comparison of the inclusive cross-sections for the
production of cumulative $\pi^+$ on Ta at  27.5 GeV extracted from the
experimental data [5] (straight line) and predictions from the
percolation scenario with ministrings (lower curve at $x=1$) and clusters 
(upper curve at $x=1$) as 
emitters.}
\end{figure}

On the other hand, the cross-sections for the protons,
result two order of 
magnitudes smaller than the experimental ones, which can be described by 
the same expression (49) with $\alpha=5.7$, $\beta=2.4$ and $C=28400$ in 
the same units. So our picture definitely does not work for the 
protons, at least at the energies corresponding to the experiment [4].

Guided by these results, we can make predictions for the cumulative $\pi^+$
production at energies corresponding to RHIC and LHC.
Taking the inelastic cross-section $\sigma_{pp}$ equal to 39 and 77 mb 
at RHIC and LHC energies respectively, we find for Pb-Pb minimum 
bias collisons the corresponding values of $\eta$ equal to 2.0 and 4.0.
So, using Eq. (42), we can read our predictions for the inclusive 
$\pi^+$ cross-section per nucleon
(in mb)) directly from Figs. 3 and 5 multiplying $d\mu/dy$ at $\eta=2$ by 
33 for RHIC and at $\eta=4$ by 43 for LHC.  

One  sees, that up to certain maximal $x$ ($\sim$ 1.6 at RHIC energy)  the 
cross-sections with clusters 
as emitters are substantially higher than with ministrings. The difference 
reaches two orders of magnitude at $1.5<x<2.5$ for LHC energy. Beyond
this maximal $x$ the cross-section with clusters abruptly goes to zero.
Thus the study of cumulative pion production at higher energies and
atomic number of participants can well distinguish between these two
mechanisms. 
\section{Discussion}
We have studied the cumulative particle production
due to the interaction of colour strings stretched between the partons
in the colliding hadrons and nuclei. The results obtained by
Monte-Carlo calculations show that 
the production rate strongly depends
on the chosen model of the string interaction. The rate is maximal and 
falls with $x$ with the minimal slope in the
string fusion  scenario in which strings fuse into the same 
form and transverse area, which corresponds to the maximal interaction 
between strings. In the percolation scenario, allowing for partial overlaps,
the rate is considerably smaller. This is, of course, to be expected, since
fusing string in the same area allows to raise the momentum much more 
effectively than with partial overlapping. The slope found in the 
percolation scenario agrees well with the experimental data at moderate
energies.

The absolute magnitudes of the found production rates also agree well 
with the existing experimental data for pions. However for the 
protons the obtained rates are far too small. This testifies that 
cumulative protons are mostly produced via a mechanism different
from colour string interactions. The fully developed Monte-Carlo simulations 
with only two strings fused, which were performed in [3], indicate that this 
mechanism is related to
the interaction between nucleons as a whole. This corresponds to
taking into account
a part of the colour states of the fused string in which a certain number of 
quark triplets are in a colourless state. In the approach employed in the
present
calculations the colour summation is done on the average and such states are
neglected (more or less in agreement with the large number of colours limit,
which lies at the basis of colour string models [6])
 
The inclusive cross-sections for cumulative production are
found to grow with energy mostly due to the growth of the
proton-proton inelastic cross-section $\sigma_{pp}$.  
They also grow with the atomic number of the participants.  Both effects 
result in the growth of the percolation parameter $\eta$.
At the RHIC and LHC energies 
predictions with ministrings and clusters
are very different.  With ministrings the cross-sectionms continue to
behave as at moderate energies with practically the same slope.
With clusters the cross-sections are much higher and fall very slowly
with $x$ (with $\alpha\sim 1.5\div 2$) up to a certain maximal $x$ after
which they abruptly fall. 
This difference in absolute magnitude and $x$- behaviour opens a way to 
distinguish between the two percolation mechanisms by experiment.

\section*{Acknowledgments}
The authors  are greatly   indebted to Yu.Shabelski
for fruitful discussions and A.Ramallo for his help in preparing the
manuscript.
M.A.B. acknowledges the financial support by the Secretaria de Estado de
Educacion y Universidades de Espanna and also of the grant
RFFI (Russia) 01-02-17137.

\section{References}

\hspace*{0.5 cm} 1. A.M.Baldin et al, Yad. Phys. 20 (1974) 1210;
Sov. j. Nucl. Phys.{\bf 20} (1975) 629.

M.I.Strikman an L.L.Frankfurt, Phys. Rep.{\bf 76} (1981) 215.

A.V.Efremov, A.B.Kaidalov, V.T.Kim, G.I.Lykasov and N.V.Slavin 
Sov. J. Nucl. Phys. {\bf 47} (1988) 868

M.A.Braun an V.V.Vechernin Nucl. Phys. {\bf B 427} (1994) 614

2.. M.A.Braun and C.Pajares, Phys. Lett. {\bf B 287} (1992) 154;
Nucl. Phys. {\bf B 390} (1993) 542,549.

 N.Amelin, M.A.Braun and C.Pajares, Phys. Lett. {\bf 287} (1992) 312;
Z.Phys. {\bf C 63} (1994) 507.

 N.Armesto, M.A.Braun, E.G.Ferreiro and C.Pajares, Phys. Rev. Lett.
{\bf 77} (1996) 3736.

3. N.Armesto, M.A.Braun, E.G.Ferreiro, C.Pajares and Yu. M.Shabelski
Astroparticle Physics {\bf 6} (1997) 329

4. Y.D.Bayukov et al. Phys. Rev. {\bf C 20} (1979) 764

5. N.A.Nikiforov et al. Phys. Rev. {\bf C 22} (1980) 700.

6. A.Capella, U.P.Sukhatme, C.-I.Tan and J.Tran Thanh Van.
 Phys. Rep. {\bf 236} (1994) 225.

7. T.S.Biro, H.B.Nielsen and J.Knoll, Nucl. Phys. {\bf B 245} (1984) 449.

8. M.A.Braun and C.Pajares, Eur. Phys. J {\bf C 16} (2000) 349.

9. M.A.Braun, F. del Moral and C.Pajares, 
Eur. Phys. J {\bf C 21} (2001) 557.

10. N.S.Amelin, N.Armesto, C.Pajares and D.Sousa, hep-ph/0103060,
to be published in Eur. Phys. J {\bf C} (2001).

11. M.A.Braun, C.Pajares and J.Ranft, Int. J. Mod. Phys. {\bf A 14} (1999) 
2689.

\end{document}